\documentstyle[12pt,aasms4,flushrt]{article}

\newcommand{\ctan}{$^{13}$C($\alpha$,$n$)$^{16}$O}
\newcommand{\ctanb}{$^{13}$C($\alpha$,$n$)$^{16}$O~}
\newcommand{\nean}{$^{22}$Ne($\alpha$,$n$)$^{25}$Mg}
\newcommand{\neanb}{$^{22}$Ne($\alpha$,$n$)$^{25}$Mg~}

\newcommand{\msb}{$M_{\odot}$~}
\newcommand{\ms}{$M_{\odot}$}
\newcommand{\cd}{$^{12}$C}
\newcommand{\ct}{$^{13}$C}

\newcommand{\ctb}{$^{13}$C~}
\newcommand{\neonb}{$^{22}$Ne~}

\begin{document}

\title{Galactic chemical evolution of heavy elements: \\
from Barium to Europium}

\author{Claudia Travaglio\altaffilmark{1}}
\affil{1. Dipartimento di Astronomia e Scienza dello Spazio, Largo
E. Fermi 5, I-50125 Firenze, Italy}
 
\author{Daniele Galli\altaffilmark{2}}
\affil{2. Osservatorio Astrofisico di Arcetri, Largo E. Fermi 5, I-50125
Firenze, Italy}
 
\author{Roberto Gallino\altaffilmark{3}}
\affil{3. Dipartimento di Fisica Generale, Universit\`a di Torino, Via
P.Giuria 1, I-10125 Torino, Italy}
 
\author{Maurizio Busso\altaffilmark{4}}
\affil{4. Osservatorio Astronomico di Torino, Strada Osservatorio 20,
I-10025 Torino, Italy}
 
\author{Federico Ferrini\altaffilmark{5}}
\affil{5. Dipartimento di Fisica, Sezione di Astronomia, Universit\`a
di Pisa, Piazza Torricelli 2, I-56100 Pisa, Italy}
 
\author{Oscar Straniero\altaffilmark{6}}
\affil{6. Osservatorio Astronomico di Collurania, I-64100 Teramo, Italy}
 
\begin{abstract}

We follow the chemical evolution of the Galaxy for elements from Ba to
Eu, using an evolutionary model suitable to reproduce a large set of
Galactic (local and non local) and extragalactic constraints.  Input
stellar yields for neutron-rich nuclei have been separated into their
$s$-process and $r$-process components. The production of $s$-process
elements in thermally pulsing asymptotic giant branch stars of low mass
proceeds from the combined operation of two neutron sources:  the
dominant reaction \ctan, which releases neutrons in radiative
conditions during the interpulse phase, and the reaction \nean,
marginally activated during thermal instabilities. The resulting
$s$-process distribution is strongly dependent on the stellar
metallicity. For the {\it standard} model discussed in this paper, it
shows a sharp production of the Ba-peak elements around $Z\simeq
Z_\odot/4$.  Concerning the $r$-process yields, we assume that the
production of $r$-nuclei is a primary process occurring in stars near
the lowest mass limit for Type II supernova progenitors.  The
$r$-contribution to each nucleus is computed as the difference between
its solar abundance and its $s$-contribution given by the Galactic
chemical evolution model at the epoch of the solar system formation. We
compare our results with spectroscopic abundances of elements from Ba
to Eu at various metallicities (mainly from F and G stars) showing that
the observed trends can be understood in the light of the present
knowledge of neutron capture nucleosynthesis. Finally, we discuss a
number of emerging features that deserve further scrutiny.

\end{abstract}


\section{Introduction}

A quantitative understanding of the Galactic evolution of nuclei
heavier than iron has been so far a challenging problem. Although some
of the basic tools of this task have been presented several years ago,
both from an observational (Spite \& Spite~1978, 1979; Sneden \&
Parthasaraty 1983; Gilroy et al.~1988) and a theoretical point of view
(Truran~1981), only recently the observational data have grown
sufficiently in number and precision to allow a direct comparison with
theoretical predictions (see McWilliam~1995 for a review).  In order to
reconstruct the evolutionary history of neutron-rich elements in the
Galaxy one has to disentangle the $s$- and $r$-contributions of all
their isotopes and follow their abundance as a function of Galactic
age. The results must then be compared with the available spectroscopic
observations of stars at different metallicities, both in the Galactic
halo and disk, mainly from F and G dwarfs and giants whose
surface abundances reflect the composition of the gas from which they formed
(Gratton \& Sneden~1994; Edvardsson et al.~1993).

Already in the seminal work by Burbidge et al.~(1957), the synthesis of
neutron-rich nuclei was attributed to two different mechanisms. The
first (the $s$-process) is characterized by {\it slow} neutron captures
and occurs mainly during hydrostatic He-burning phases of stellar
evolution. Here {\it slow} means that most unstable nuclei encountered
along the neutron capture path have time to decay before capturing
another neutron. The second mechanism (the $r$-process) is due to the
more {\it rapid} neutron captures and is generally associated with
explosive conditions in supernovae (hereafter SNe). For a comprehensive
review on the astrophysical sites of neutron capture nucleosynthesis,
see Wheeler, Sneden, \& Truran (1989).

The current understanding of the $s$-process is supported by many
observational and theoretical works (summarized in Sect.~2), indicating
that the heavier $s$-nuclei, from Sr to Pb, belonging to the so-called
{\em main} component (Clayton \& Ward 1974; K\"appeler et al. 1982),
are synthesized during the thermally pulsing asymptotic giant branch
(TP-AGB) phase of low-mass stars, mainly in the mass range $\sim
1$--4~\msb (see e.g.  Gallino, Busso, \& Lugaro 1997;
Busso, Gallino, \& Wasserburg 1999, hereafter BGW).  A different site
for the production of $s$-process nuclei is required to account for the
{\em weak} component, i.e. for $s$-nuclei below the Sr-peak. This site
was identified in the advanced evolutionary phases of massive stars
(Lamb et al. 1977; Prantzos et al. 1990; Raiteri et al.~1993).

As for the $r$-process, the actual astrophysical environment is still a
matter of debate (Woosley et al.~1994; Wheeler, Cowan, \&
Hillebrandt~1998).  It has been recently shown that at least two
different supernova sources are required for the synthesis of the
$r$-nuclei below or beyond the neutron magic number $N = 82$
(Wasserburg, Busso, \& Gallino~1996; see also Sneden et al.~1998).  For the
elements considered here, we make the simplifying hypothesis that their
$r$-contributions in the solar system can be considered of primary
origin (i.e. with yields independent of the initial stellar
metallicity). Hence, after the $s$-contributions at the epoch of the
solar formation have been estimated with the use of a reliable model
for the chemical evolution of the Galaxy, the $r$-process abundance
fractions can be derived by subtracting the $s$-fractions from the
solar abundances.

The study of the Galactic evolution of elements produced by $s$- and
$r$-processes was addressed in previous investigations only at a
semi-quantitative level. For example, Andreani, Vangioni-Flam, \&
Audouze~(1988), and, more recently, Pagel \&
Tautvai\v{s}ien\.{e}~(1997) have assumed {\em ad hoc} stellar yields
and time delays for representative $s$ and $r$ elements in order to fit
the observational data.  With a different approach, Mathews, Bazan, \&
Cowan~(1992), following a preliminary analysis by Clayton (1988),
examined the effects introduced by the primary nature of the \ctb
neutron source on the $s$-process yields, and inferred that for
metallicities $Z>10^{-3}Z_\odot$ the neutron exposure $\tau \equiv \int
n_n v_T dt$ (where $n_n$ is the neutron density and $v_T$ the thermal
velocity) roughly scales as $Z^{-1}$. Under the crude assumption that
the production (in mass fraction) of a given $s$-isotope depends on the
product of the neutron exposure times the abundance of iron group
seeds, Mathews et al.~(1992) estimated the stellar $s$ yields by simply
scaling the solar $s$-abundances with metallicity.  Today the work of
Mathews et al.~(1992) needs to be revised, as quantitative yields are
now available from detailed nucleosynthesis calculations based on TP-AGB models
for different masses and metallicities, with updated reaction
networks and neutron capture cross sections.  In addition, as
anticipated by Truran (1981), at low metallicities the $r$-process
contribution to the various elements becomes dominant.

In this paper, we concentrate on elements from Ba to Eu, not only
because they belong to a major abundance peak, but also because they
contain species of very different origin, ranging from almost pure
$r$-process production, as is the case  for Eu, to a dominant
$s$-process origin, as is the case for Ba. The paper is organized as
follows:  in Sect.~2 we analyze the neutron capture process in AGB
stars; in Sect.~3 we describe the adopted model for the chemical
evolution of the Galaxy and the updates introduced in the present
work; in Sect.~4 we show the model results for the elements considered
(Ba, La, Ce, Pr, Nd, Sm, and Eu) and for their isotopes; finally, in
Sect.~5 we summarize the main conclusions and point out a few aspects
deserving further analysis.

\section{The production of $s$-process nuclei in Asymptotic Giant Branch stars}

\subsection{Previous work}

The first quantitative model for the production of $s$-process nuclei
with $A > 85$ was presented by Truran \& Iben (1977) and by Iben \&
Truran (1978). It was based on the activation of the \neanb neutron
source in TP-AGB stars of intermediate mass ($5\lesssim M/M_\odot
\lesssim 8$).  However, subsequent reanalyses showed that the attempts
to reproduce the solar main component through the \neanb neutron source
suffer from major nuclear problems (Despain 1980; Howard et al. 1986;
Mathews et al. 1986; Busso et al. 1988). On the other hand,
observations revealed that the luminosities of MS, S and C (N type)
stars are generally much lower than predicted by the above models, thus
pointing out that the dominant contributors to the synthesis of carbon
and $s$-process elements are AGB stars of lower mass (see e.g.  Clegg,
Lambert, \& Bell~1979; Blanco, McCarthy, \& Blanco~1980; Smith \&
Lambert~1986; Lambert et al.~1991). In low-mass AGB stars the maximum
temperature reached during thermal instabilties in the He shell is too
low ($T_{\rm max}\simeq 3\times 10^8$~K) for \neonb to be significantly
consumed.  Consequently, neutron captures must be driven by the
alternative \ctanb reaction (Cameron 1954), which can easily be
activated at much lower temperatures (for the rate of this reaction see 
Caughlan \& Fowler 1988 and Denker et al. 1995).

The amount of \ctb left in the He intershell from previous H burning is
by far too low to be of any relevance in the neutron production.  A
challenging problem is therefore to explain how a sufficient amount of
\ctb can be built locally. The common hypothesis is that some protons
can penetrate below the formal convective border of the envelope into
the top layers of the He intershell.  At H re-ignition, these protons
are captured by the abundant $^{12}$C left by partial He burning in the
previous convective pulse, giving rise to a tiny layer enriched in
$^{13}$C (the so-called $^{13}$C pocket).  The first mechanism
considered in order to generate a $^{13}$C pocket was the
semi-convective mixing of H into the $^{12}$C-rich and He-rich
intershell, driven by increased carbon opacity in the cool expanding
envelope after the quenching of a thermal pulse (Iben \&
Renzini~1982a,b). This process was shown to be relevant for Population
II stars (Hollowell \& Iben~1988, 1989).  Once formed, the \ctb pocket
was assumed to be engulfed by the next convective pulse releasing
neutrons. The resulting neutron density, originally estimated to be
very high ($\ge$ 10$^{11}$ n/cm$^3$, see Malaney 1986) was subsequently
shown to be of the order of (5--10)$\times 10^8$ n/cm$^3$ (K\"appeler
et al 1990; Gallino, Raiteri, \& Busso 1993).  However, semiconvection
was found to become inefficient in Galactic disk stars (Iben 1983), and
no obvious alternative from simple diffusive mixing was found to work
(Iben 1982).

The penetration of the convective envelope into the partially He-burnt
zone after each pulse (the so-called {\it third dredge-up}) leaves a
sharp H/C discontinuity between the H-rich envelope and the 
$^{12}$C-rich and He-rich intershell, which is likely to be smoothed by some
sort of mixing at the interface, by chemical diffusion, semiconvection,
or rotational shear (Langer et al.~1999).  To address this point, a
rather sophisticated treatment of the hydrodynamic behavior at the H/He
boundary would be required. First attempts with a diffusive approach
(Herwig et al.~1997), or with a fully hydrodynamical formulation
(Singh, Roxburgh, \& Chan 1998) have been recently presented and do
show formation of a \ctb reservoir.  Indeed, penetration of protons at
the top of the \cd-rich intershell appears to be plausible, but the mass
involved and the resulting H profile have still to be treated as
relatively free parameters. A dedicated  effort on hydrodynamical
grounds constitutes a major challenge for AGB calculations and
$s$-process nucleosynthesis studies.

\subsection{$s$-process yields from AGB stars: FRANEC results}

A more indirect approach is to consider different choices for
the mass and profiles of the \ctb pocket, and to estimate their effects
on the nucleosynthesis results for AGB models of various
metallicities.  The TP-AGB models by Straniero et al.~(1995, 1997)
showed that any amount of \ctb produced by a proton penetration below
the convective envelope border burns radiatively in the interpulse
period, at a relatively low temperature ($T\gtrsim 0.9\times 10^8$~K)
and neutron density ($n_n \lesssim 10^7$ n/cm$^3$), giving rise to a
tiny layer extremely enriched in $s$-process isotopes that will be
engulfed by the next convective instability.  The resulting pattern of
AGB nucleosynthesis, and its dependence on the initial metallicity of
the star, have been recently discussed by Gallino et al.~(1999).  The
changes introduced by these recent studies are important: not only the
neutron density is lower by a factor of $\sim 20$ than in previous
estimates, but the distribution of neutron exposures can no longer be
approximated by an exponential distribution, as commonly assumed
previously, with relevant consequences for the dependence of neutron
captures on metallicity (Busso et al.~1997).  A thorough analysis of
the implications of the new models is presented by BGW.

The stellar evolution results shown in this paper are all based on
calculations performed with FRANEC (Frascati Raphson-Newton
Evolutionary Code, see Chieffi \& Straniero~1989). Stellar models
appropriate for Population I stars have already been presented in
Straniero et al.~(1997) and Gallino et al.~(1998); additional models
have been computed specifically for the present paper. Our set of AGB
models includes stars of masses from 1.5, 2, 3 and 5~\ms, and
metallicity $Z=0.006$, 0.01 and 0.02. Mass-loss was taken into account
adopting the parametrization by Reimers (1975) with values of the
$\eta$ parameter ranging from 0 to 5.

The FRANEC AGB models reproduce self-consistently the third dredge-up
episode and the consequent mixing of freshly synthesized $s$-processed
material (together with $^{4}$He and $^{12}$C) to the surface of the
star.  According to Straniero et al.~(1995, 1997) and Gallino et al.
(1998), the third dredge-up starts after a limited number of thermal
pulses, increasing its efficiency up to a maximum while the pulsing
mechanism goes on, then decreases as mass loss and the advancement of
the H-burning shell reduce the envelope mass, and eventually ceases
when the latter decreases below about 0.5 \ms.

The mass and the profile of the \ctb pocket were constrained by Busso
et al. (1995, see also Busso et al.~1999) by comparing the results of
neutron capture models with spectroscopically determined abundances in
Population I AGB stars and in Population II Ba and CH stars with [Fe/H]
$\le -1$ (Smith \& Lambert~1990; Luck \& Bond 1991; Vanture 1992; Plez,
Smith, \& Lambert~1993). This was done by keeping constant the total
mass of the \ctb pocket and parametrically varying the concentration of
\ctb by factors from 0 to 2 times the profile indicated as ``standard''
(ST) by Gallino et al. (1998), corresponding for the choice of the
mass of the \ctb pocket made for this work to $3.1\times
10^{-6}$~\ms of \ct. As a result of this study, the ST choice
for the \ctb pocket, when used in an AGB stellar model of metallicity
$Z=Z_{\odot}/2$, was shown to reproduce remarkably well the abundance
distribution of the solar main component (see also Arlandini et al.
1999, in preparation). However, AGB stars in the Galactic disk show a
significant spread in their abundances of $s$-process elements (see
discussion in Busso et al.~1995 and Busso et al.~1999). This spread
can be attributed, at least in part, to variations in the \ctb
profile, since the $s$-process efficiency depends strongly and non
linearly on the local \ctb abundance. Following this indication, we
have explored different extensions in mass of the \ctb pocket, while
keeping for all metallicities the same \ctb profile as in ST case.  We
will later comment briefly on these choices.  Let us mention here that
the same choice was found to be appropriate to account for the
$s$-process isotopic signatures of heavy elements (among which Ba, Nd,
and Sm) in presolar grains most likely condensed in the circumstellar
envelopes of AGB stars of solar metallicity (Gallino et al.~1993;
Gallino et al.~1997) and recovered in meteoritic material (Ott \&
Begemann~1990; Prombo et al.~1993; Zinner~1997; Hoppe \& Ott~1997).

At the end of the TP-AGB phase, the $s$-process contributions to the
ISM are determined by the amount of matter cumulatively dredged-up from
the He intershell to the surface and lost by stellar winds. This
quantity is shown in Tab.~1 for the 2, 3, and 5~\msb stellar models and
different  metallicities (for the 5~$M_\odot$ star we did not compute
the model at $Z=0.006$ but we doubled the cumulative dredged-up mass
mixed into the envelope with respect to the solar case, in analogy with
the results obtained for the 2 and 3 \msb cases).

\begin{table}[t]
\begin{center}
TABLE 1\\
{\sc Cumulative dredged-up mass in the envelope (in \ms) for AGB stars}\\ 
\vspace{0.5em}
\begin{tabular}{llll}
\hline\hline
 $Z$  & 2 \msb  & 3 \msb & 5 \msb\\
\hline
0.006 & 0.030 & 0.078 & 0.12 \\ 
 0.02 & 0.018 & 0.044 & 0.06 \\ 
\hline
\hline
\end{tabular}
\end{center}
\end{table}

In Fig.~1 we show the abundance by mass of Ba in the He shell material
cumulatively mixed to the surface as a function of [Fe/H] for a 2~\msb
AGB model, for different assumptions on the total amount of \ctb nuclei
present in the $^{13}$C pocket. The dashed thick lines represent the
extreme values chosen in the parametrization (0.17 and  1.33 times the
ST case), whereas thin lines show intermediate cases.  The thick
continuous line represents the unweighted average yield calculated over
all the cases shown. In Fig.~2 we show the full set of unweighted
average yields for Ba, La, Ce, Nd, Pr, Sm, and Eu.

It is evident from Fig.~1 and 2 that there is a strong dependence of
$s$-process yields on the initial stellar metallicity. This important
result deserves additional comments. Since the build up of heavy nuclei
requires neutron captures starting from Fe seeds, the $s$-process is
expected to decline with declining metallicity, i.e. to be of {\it
secondary} nature.  However, the abundances produced depend not only on
the initial Fe concentration, but also on the neutron exposure.  The
concentration of $^{13}$C in the pocket is of {\em primary} nature (it
is built from H and freshly made C, hence is independent of
metallicity), while the abundance of the neutron absorber $^{56}$Fe
varies linearly with $Z$. Therefore the neutron exposure $\tau$
(roughly proportional to the ratio $^{13}$C/$^{56}$Fe) is expected to
scale roughly as $Z^{-1}$. This dependence would compensate the mentioned
secondary nature of the $s$-elements, if the yields of $s$-nuclei were
linearly dependent on both $\tau$ and $Z$.

One has however to notice that the dependence of $s$-process yields on
the neutron exposure is very complex and non linear (Clayton~1968). In
general, the production factor of a given $s$-element (e.g. Ba) depends
on the average abundances achieved by all the neutron-rich nuclei, and
varies with $Z$ also for metallicities close to solar.  As shown in
Fig.~1, in the Galactic disk, for a range of metallicities covering
about 1 dex, the strong increase of the $s$-process yields with
increasing neutron exposure dominates and the secondary nature of these
nuclei is {\it overcompensated}, so that they actually {\it increase}
for decreasing $Z$, instead of remaning constant, as the previous
simple discussion would have suggested.

At low metallicities, an important role is also played by light neutron
absorbers, which are mainly primary in nature, as stressed by Clayton
(1988) and by Mathews et al.~(1992). Among these nuclei one has to
include $^{12}$C, $^{16}$O and most of $^{22}$Ne with its progeny. This
is so because a large fraction of $^{22}$Ne ultimately derives from the
processing of the primary $^{12}$C dredged-up into the envelope from
the He-shell, which is transformed first to $^{14}$N by shell H-burning
and then to $^{18}$O and $^{22}$Ne by He-burning.  Notice that the
recently measured neutron capture cross section of one of the major
primary neutron absorbers, $^{16}$O (Igashira et al. 1995), is 170
times higher than the previous theoretical estimate quoted in Beer,
Voss, \& Winters (1992).  On the whole, the importance of the
moderating effect on the neutron density by all light neutron absorbers
cannot be discarded.

In addition, as discussed by Gallino et al.~(1998), radiative \ctb
burning builds the distribution of $s$-elements in steps of relatively
large $n$-exposition. This gives rise to higher values of the Ba/Zr
ratio than feasible in any process working inside the pulses (that
would produce simpler exponential $n$-exposure distributions).  This
last feature is a necessary condition for reproducing the abundance
observations in CH giants near [Fe/H] $= -1$ (Vanture 1992; Busso et
al.~1999) and also helps in maintaining a high production of Ba over a
wide range of metallicities.

The trends shown in Fig.~2 can be understood through the interplay of the
phenomena outlined above.  Starting from AGB stars of nearly solar
metallicity, first the $s$-fluence builds up the $s$-elements belonging
to the Zr-peak, at neutron magic $N$ $=$ 50. Then the Zr-peak yields
decrease while the Ba-peak production increases as shown in the Figure,
reaching a maximum at [Fe/H] $\simeq -0.6$.  For lower metal contents,
the $n$-flux skips Ba and feeds Pb, which reaches a maximum production
yield at [Fe/H] $= -1$ (Gallino et al. 1999). Eventually, also Pb
decreases making the secondary nature of the $s$-process evident.  At
these very low metallicities, essentially all the Fe group seeds are
converted to Pb.  Also thanks to the additional effect of light
(primary) neutron absorbers, below [Fe/H] $\simeq -2$ only a very small
$s$-signature is present on heavy elements.

\section{Galactic chemical evolution model}

Our model for the chemical evolution of the Galaxy is described in
detail by Ferrini \& Galli (1988), Galli \& Ferrini (1989) and Ferrini
et al. (1992). The model has been adopted to investigate different
aspects of the chemical evolution of the Galaxy as well as global
properties of external galaxies (see e.g. Moll\`a, Ferrini, \&
D\'{\i}az~1997 and refererences therein). Here we summarize its
physical characteristics relevant to our study and briefly discuss the
modifications to the original nucleosynthetic prescriptions introduced
in the present work.

The Galaxy is divided into three zones, halo, thick disk, and thin
disk, whose composition of stars, gas (atomic and molecular) and
stellar remnants is computed as function of time up to the present
epoch $t_{\rm Gal}=13$~Gyr. Stars are born with an initial composition
equal to that of the gas from which they formed.  The formation of the
Sun takes place 4.5~Gyr ago, i.e. at epoch $t_\odot=8.5$~Gyr. The thin
disk is divided into concentric annuli, with no radial flow allowed,
and is formed from material infalling from the thick disk and the
halo.  In the present work we neglect any dependence on galactocentric
radius in the model results as well as in the observational data (see
Moll\'a et al.~1997 for an application of the model to the study of
radial abundance gradients), and we concentrate on the evolution inside
the solar annulus, located 8.5~kpc from the Galactic center.

The star formation rate $\psi(t)$ is not assumed {\em a priori}, but
is obtained as the outcome of self-regulating processes occurring in the
molecular gas phase, either spontaneous or stimulated by the presence
of other stars.  The nucleosynthesis prescriptions are based upon the
classic ``matrix formalism'' first introduced by Talbot \& Arnett
(1973).  The mass of the element $i$ ejected by a star of mass $M$
after a time $\tau(M)$ from birth is assumed to be a linear combination
of the masses of elements $j$ initially present in the star:
\begin{equation}
M_i^{\rm ej}=\sum_jQ_{ij}M_j^{\rm in},
\end{equation}
where the matrix elements $Q_{ij}$ depend on the mass and the 
metallicity of the star.

In the framework of Galactic evolution, it is convenient to consider
separately the contribution of stars in different mass ranges. The
model follows the evolution of ({\em i}\/) single low- and
intermediate-mass stars ($0.8~M_\odot\leq M \leq M_\star$) ending their
life as He or C-O white dwarf, ({\em ii}\/) binary systems able to
produce Type {\small I} supernovae, and ({\em iii}\/) single massive
stars ($M_\star\leq M\leq 100~M_\odot$), the progenitor of Type {\small
II} supernovae. The value of $M_\star$ depends on metallicity; we
assume $M_\star=6$~\ms\/ for $Z\leq 10^{-3}$ and $M_\star=8$~\ms\/
otherwise (see for references Tornamb\`e \& Chieffi~1986).  For each
stellar group, the restitution rate $W_{i,k}(t)$ of element $i$ is
given by
\begin{equation}
\label{wpn}
W_{i,k}(t)=\int_{M_{\rm low}}^{M_{\rm upp}}\sum_j Q_{ij}[M,Z(t)]X_j
[t-\tau(M)]\psi[t-\tau(M)]\phi(M)dM,
\end{equation}
where the index $k$ refers to the zone of the Galaxy, e.g. halo, thick
disk and thin disk, and $X_j[t-\tau(M)]$ is the abundance (by mass) of
element $j$ in the ISM at the time of birth of a star of mass $M$. The
adopted initial mass function $\phi(M)$ is discussed in Ferrini et
al.~(1992).

We have updated the nucleosynthesis prescriptions of Ferrini et
al.~(1992) for SNII and SNI adopting the yields computed by Woosley \&
Weaver (1995) and Thielemann, Nomoto, \& Hashimoto (1996),
respectively.  As an example of the model results, Fig.~3a shows the
run of star formation rate vs. [Fe/H] in the three Galactic zones for
our Standard Model (corresponding to Model B of Pardi, Ferrini, \&
Matteucci~1995). The [Fe/H] scale here is indicative of a time scale,
albeit a non-linear one. The corresponding values of [O/Fe] vs. [Fe/H]
for the three zones, computed with the revised yields, are plotted in
Fig.~3b, together with an updated compilation of observational data. We
have included abundance determinations obtained from measurements of
the [OI] forbidden lines (Gratton \& Ortolani~1986; Barbuy~1988; Barbuy
\& Erdelyi-Mendes~1989; Sneden et al.~1991; Spite \& Spite~1991;
Edvardsson et al.~1993) and from the OH bands in the near UV (Israelian
et al.~1998).

We see from Fig.~3a-b that the halo phase lasts approximately up to
[Fe/H] $\lesssim -1.5$, the thick-disk phase covers the interval
$-2.5\lesssim {\rm [Fe/H]} \lesssim -1$, and the thin-disk phase starts
at approximately [Fe/H] $\gtrsim -1.5$. The slow decline of [O/Fe] for
increasing [Fe/H], due to contribution of SNI as the evolution
proceeds, is reproduced by the model over more than three decades in
metallicity.

Recently Israelian, L\'opez, \& Rebolo (1998) have measured oxygen
abundances from the OH bands in the near UV in a sample of stars in the
metallicity range $-3.0<$[Fe/H]$<0.3$ (also included in Fig.~3b).  They
found values significantly higher than those reported by other authors
shown in the same Figure, with a continuous increase of [O/Fe] with
decreasing [Fe/H]. To reproduce these high values, one possibility is
to increase the O contribution with respect to Fe of the most massive
Type II SNe by a factor $\sim 2$.  This modification, of course, has no
consequence on the chemical evolution of the $s$-process elements.

\section{Results for the Galactic evolution of $s$- and $r$-elements}

In this Section we present our results for the evolution of Ba, La, Ce,
Nd, Pr, Sm and Eu in the Galaxy, by considering separately the $s$- and
$r$-contributions.  Then we compute the abundance of these elements
resulting from the sum of the two processes and we compare  model
results with the available spectroscopic observations of field stars 
at different metallicities.

\subsection{Galactic $s$-process evolution}

The $s$-process yields discussed in Sect.~2 and shown in Fig.~2 allow
us to estimate the chemical enrichment of the Galaxy at different
times.  We show in Fig. 4 the resulting Ba $s$-fraction in the thin
disk, compared with spectroscopic abundances obtained from various
observational campaigns (Gratton \& Sneden 1994; Woolf, Tomkin, \&
Lambert 1995; Fran\c cois 1996; McWilliam et al. 1998; Nissen \&
Schuster~1997; Norris, Ryan, \& Beers~1997; Jehin et al.~1998;
Mashonkina, Gehren, \& Bikmaev~1999).

Though we have considered stars from 2 to 8 \msb, the mass range from
which we get the dominant production of Ba is $2 - 4$~\msb (as shown in
Fig. 4), well representative of the most common low-mass AGB stars.
More massive AGBs, in the range $4 - 8$~\ms, do not give a relevant
contribution to Ba-peak elements (see the following discussion).

As anticipated in Sect.~2, the direct comparisons of spectroscopic
observations of AGB stars, solar system abundance determinations,
isotopic anomalies in presolar SiC grains, with the predictions of
stellar models constrain the mass of the $^{13}$C pocket to within a
certain range around a standard value.  In the framework of Galactic
chemical evolution, we adopt {\em averaged} stellar yields (shown in
Fig.~1 and Fig.~2), calculated by assuming that all values of the
$^{13}$C pocket are equally probable within this interval. However, to
test the sensitivity of our results to the choice of this fundamental
parameter, we have also computed the Galactic evolution of the
$s$-elements with stellar yields corresponding to the lower and upper
bounds for the mass of the $^{13}$C pocket. These aspects will be
further discussed at the end of this subsection.

\begin{table}[t]
\begin{center}
TABLE 2\\
{\sc $s$-process fractional contributions (\%) at $t=t_{\odot}$ \\
with respect to solar system abundances}\\

\vspace{0.5em}
\textheight 25cm
\textwidth 18cm
\voffset -2.5cm
\begin{tabular}{lllll}
\hline\hline
  & case I$^{(\rm a)}$ & case II$^{(\rm b)}$ & case III$^{(\rm c)}$ & KBW89$^{(\rm d)}$ \\
\hline
Ba  & 10 & 253 & {\bf 80} & 88 \\
La  & 7 & 200  & {\bf 61} & 75 \\
Ce  & 8 & 248  & {\bf 75} & 77 \\
Pr  & 5 & 156  & {\bf 47} & 45 \\
Nd  & 5 & 178  & {\bf 54} & 46 \\
Sm  & 3 & 99   & {\bf 30} & 30 \\
Eu  & 0.6 & 20 & {\bf 6}  & 3  \\
\hline\hline
\end{tabular}
\end{center}
\small{
\hspace{9em}$^{(\rm a)}$ -- $^{13}$C concentration: 0.17 times the ST case
\vspace{-1em}

\hspace{9em}$^{(\rm b)}$ -- $^{13}$C concentration: 1.33 times the ST case
\vspace{-1em}

\hspace{9em}$^{(\rm c)}$ -- $^{13}$C pocket mass averaged
\vspace{-1em}

\hspace{9em}$^{(\rm d)}$ -- K\"appeler, Beer, \& Wisshak (1989)
}
\end{table}

In Table~2 we list the resulting $s$-fractions (with respect to the
corresponding solar abundances) at $t=t_\odot$. In the first two
columns we show the results obtained using a mass of the $^{13}$C
pocket scaled by 0.17 (case I) and by 1.33 (case II) from the ST case.
In the third column we show the results for the average yields (case
III). The corresponding stellar yields are shown in Fig.~2 (case III),
and, for Ba only, in Fig.~1 (case I, II and III). We also compare, in
the fourth column, our results with the previous estimates by
K\"appeler, Beer, \& Wisshak~(1989), obtained with the so called {\it
classical analysis}, based on model-independent analytical expressions
(an exponential form) for the neutron irradiation.  Notice that the
results obtained with the extreme values of the mass of $^{13}$C are in
strong disagreement with either the results of the classical analysis
or the observed solar abundances.  As an illustration, we see from
Table~2 that at $t = t_\odot$ the Ba $s$-fraction obtained with the
average yields is 80 \% of the solar Ba abundance, whereas the lower
and upper values of the mass of $^{13}$C give a Ba $s$-fraction of 10
\% and 253 \%, respectively. Notice also that the average yields
give results in reasonable agreement with the classical analysis.

It is evident from Fig.~4 that the $s$-process contribution dominates
the Galactic evolution of Ba starting from [Fe/H] $\simeq - 1.5$.  At
lower values of [Fe/H], independently on the characteristics of the
chemical evolution model, the contribution of $s$-process
nucleosynthesis rapidly decreases due to the strong dependence of
stellar yields on metallicity:  {\em hence the contribution of low-mass
AGB stars is far too low to account for the observed abundances in the
range} $-3\lesssim$ [Fe/H] $\lesssim -1.5$.  In order to check the
sensitivity of this result on model inputs, we have considered the
effects of the contribution of intermediate mass stars in the range
4--8~\ms, adopting the yields of a 5~\msb AGB model with solar
metallicity (Vaglio et al.~1999) as representative of this mass
interval. These stars activate the $^{22}$Ne($\alpha$,$n$)$^{25}$Mg
reaction during their TP-AGB phase more efficiently than in low-mass
stars, because of the higher temperature reached at the bottom of the
convective pulse ($T_{\rm max}\lesssim 3.5\times 10^8$~K). In contrast,
the formation of a \ctb pocket is less certain, due to the reduced mass
of the He intershell (by one order of magnitude). In this mass range
the observational constraints are scarce and indirect: no AGB star of
the Galaxy, for which we have a spectroscopic estimate of abundances,
can be unambiguously attributed to this mass range, though in some
post-AGB stars there is some suggestion that an intermediate mass star
may have been the progenitor (Busso et al.~1999). Given such a lack of
constraints, for these stars we have computed $s$-process
nucleosynthesis induced by the $^{22}$Ne($\alpha$,$n$)$^{25}$Mg neutron
source alone.  We find that the neutron flux is too low for these stars
to significantly contribute to the Galactic chemical evolution of the
Ba-peak elements, whereas there is a consistent production of the
elements belonging to the first $s$-peak:  Sr, Y, and Zr. In
particular, the contribution of intermediate mass stars only increase
the Ba abundance at $t=t_\odot$ by about 3\%, and let the rising of
[Ba/Fe] vs. [Fe/H] shift at slightly lower metallicity, due to the
shorter stellar lifetimes.

The results of the Galactic chemical evolution are not particularly sensitive
to the value of the mass contained in the \ctb pocket adopted. For example,
various tests for the $s$-process nucleosynthesis have been made by 
progressively reducing the \ctb mass giving results very close 
to the ones illustrated in Fig.~1. This is inherent to the
$s$-process nucleosynthesis mechanism, which, increasing the number of
available neutrons per Fe group seed (i.e. with decreasing
metallicity), proceeds through the accumulation of neutron-magic
nuclei, at $N= 50$, 82, and 126, at the Zr-peak, Ba-peak and at Pb,
respectively.  Varying the total mass involved in the \ctb pocket in
the present calculations by a factor of 2 is equivalent to vary by only
30 \% the final yield, because of the overlapping mechanisms between
subsequent pulses.  This is an indication to extend 
the analysis to cases with somewhat higher abundances of \ctb
nuclei with respect to the upper case ($1.3 \times$~ST) considered in
the present calculations, as suggested by the comparison of
spectroscopic abundances of some $s$-enhanced stars (Busso et
al.~1999). 

Despite the large number of approximations (the total
amount of dredged-up material, which depends on the stellar
evolutionary code and on the adopted law of mass loss by stellar winds,
the parametrization on the \ctb pocket, etc.) the present results
appear to reasonably reproduce the $s$-process distribution of the
heavy elements, from the Ba-peak up to Pb (see also Gallino et al.
1998, 1999).

\subsection{Galactic $r$-process contribution}

As first suggested by Truran~(1981), the presence of $r$-process
elements in low metallicity halo stars (Gratton \& Sneden~1994;
McWilliam et al.~1995; McWilliam~1998) is indicative of a prompt
enrichment of the Galaxy in these elements, possibly by early
generations of massive stars.

It is possible to constrain quantitatively the enrichment of
$r$-process elements in the ISM during the evolution of the Galaxy on
the basis of the results presented in Sect.~4.1 and in Table~2 for the
$s$-process contribution at $t=t_\odot$.  The so-called $r$-process
residuals are obtained by subtracting the $s$-process contribution
$N_s/N_\odot$ from the fractional abundances in the solar system taken
from Anders \& Grevesse~(1989):
\begin{equation}
     N_r/N_\odot = (N_\odot - N_s)/N_\odot. 
\end{equation}
In the case of Ba we obtain a $r$--residual of 20\%. 
The assumption that the $r$-process is of primary nature 
and originates from massive stars allows us to estimate the 
contribution of this process during the evolution of the Galaxy.
In the case of Ba, for example, we have
\begin{equation}
   \left(\frac{{\rm Ba}}{{\rm O}}\right)_{r,\odot}
   \simeq 0.2\left(\frac{{\rm Ba}}{{\rm O}}\right)_\odot.
\end{equation}
Since the $s$-process does not contribute at low metallicity (see 
Sect.~4.1 and Fig.~4), for Population II stars we have approximately
\begin{equation}
   \left(\frac{{\rm Ba}}{{\rm O}}\right)
   \simeq \left(\frac{{\rm Ba}}{{\rm O}}\right)_{r,\odot},
\end{equation}
which yields 
\begin{equation}
   \left[\frac{{\rm Ba}}{{\rm Fe}}\right]=
   \left[\frac{{\rm Ba}}{{\rm O}}\right]+\left[\frac{{\rm O}}{{\rm
Fe}}\right]
   \simeq \log(0.2) + 0.6 \simeq -0.1\;{\rm  dex},
\end{equation}
assuming a typical [O/Fe] $\simeq 0.6$~dex for Population II stars.
Thus, the $r$-process contribution for [Fe/H] $\lesssim -1.5$ dominates
over the $s$-contribution and roughly reproduces the observed values.
The simple argument presented here ignores, however, the effects of the
finite lifetimes of stars, the distribution of stellar masses, and the
time dependence of the star formation rate.  All these effects are
instead taken into account in our chemical evolution model. The results
shown in Fig.~5 for Ba and Eu confirm the estimate given above, and
provide a strong constraint on the mass range of SNII contributing to
$r$-process nucleosynthesis.  In fact, the characteristic decline of
[$r$/Fe] as function of [Fe/H] suggested by observations of metal-poor
stars (Cowan, Thielemann, \& Truran~1991; Mathews et al. 1992; Gratton
\& Sneden~1994) can be naturally explained by a time delay between the
O-rich (and partly Fe-rich) material ejected by the more massive SNII
($M\ge 15$~\ms), and $r$-process material ejected by the lower mass
SNe. This delay may reflect the actual difference in stellar lifetimes,
but can be amplified by non-instantaneous mixing processes in the ISM.
As for the scatter of spectroscopic data at very low
metallicities, this problem will be discussed in the next subsection.

In particular we have considered in Fig.~5 the following cases: (a) the
full range of SNII, (b) SNII in the mass interval $10 \le M/M_\odot \le
12$, and (c) SNII in the mass interval $8 \le M/M_\odot \le 10$.  It is
evident that in order to reproduce the typical increasing trend of
[Ba/Fe] or [Eu/Fe] at [Fe/H] $\lesssim -1.5$, the production of Fe at
low metallicity must have occurred substantially before the production
of the $r$-process component of Ba and Eu.  According to our model,
SNII in the mass range $8 \le M/M_\odot \le 10$ appear to be good
candidates for the {\em primary} production of $r$-nuclei, whereas a
range extended to much higher masses seems to give results in conflict
with the available observations.  We note in addition that the
identification of low-mass SNe as a possible site for the $r$-process is
also supported by recent theoretical SN models by Wheeler et
al.~(1998), Freiburghaus et al.~(1998), and Meyer \& Brown~(1997).

\subsection{$s$- and $r$-process contributions to Galactic chemical
evolution}

In this Section we present our results for the Galactic chemical
evolution of elements from Ba to Eu, based on the assumptions discussed
in Sect.~$4.1 - 4.2$, namely that the $s$-process contribution to the
production of these elements comes from $2 - 4$~\msb AGB stars, and the
$r$-process contribution originates from SN in the range $8 - 10$~\ms.

Fig.~6 to 11 show the Galactic evolution of Ba, La, Ce, Nd, Pr, Sm,
computed by adding the $s$- and $r$-process contributions (abundances
are given with respect to Fe and Eu in the upper and lower panels
respectively).  Fig. 12 shows the resulting evolution of [Eu/Fe] as a
function of [Fe/H], compared with spectroscopic abundances determined in 
dwarf and giant stars of different metallicities.

The [element/Fe] ratios (upper panels of Fig. $6 - 11$) provide
information about the enrichment relative to Fe in the three Galactic
zones, making clear that a delay in the $r$-process production with
respect to Fe is needed in order to match the spectroscopic data in the
range $-3\lesssim$ [Fe/H] $\lesssim -2$.  The observations show that
the [element/Fe] ratios begin to decline in metal-poor stars, with a
sharp drop at [Fe/H] $\lesssim -2.5$.  This trend can be naturally
explained by the finite lifetimes of stars at the lower end of the
adopted mass range: massive stars in the early times of the evolution
of the Galaxy evolve quickly, ending as SNII producing O and Fe. Later,
less massive stars explode as SNII, producing $r$-process elements
and causing the sudden increase in [element/Fe].

Since Eu is mostly produced by $r$-process nucleosynthesis (94\% at
$t=t_\odot$, see Tab. 2), the [element/Eu] abundance ratios (bottom
panels) provide a direct way to judge the relative importance of the
$s$ and $r$ channels during the evolution of the Galaxy. According to
our model, at low metallicity the $r$-process contribution is dominant,
and the [element/Eu] ratio is approximately given by the element's
$r$-fraction computed with the $r$-residuals discussed in Sect.~4.2.
On the contrary, for [Fe/H] $\gtrsim -1.5$, the $s$-process
contribution takes over, and the [element/Eu] ratios rapidly increase
approaching the solar values.  This trend is particularly evident in
the observational data for Ba and Ce, as these elements have the
largest $s$-process contribution.

The agreement of the model results with observational data for
$s$-process dominated elements (e.g. Ba and Ce) provides a strong
support to the validity of the nucleosynthesis prescriptions adopted in
this work and AGB modelings with the FRANEC code, including the
estimate of the $s$-process and \cd-rich material dredged-up in AGB
envelopes and eventually mixed with the ISM by stellar winds.  In
turn, the $s$-fractions obtained from our model at $t=t_\odot$, listed
in Table~2, are used to normalize the $r$-process nucleosynthesis and
therefore determine the element abundances at earlier Galactic epochs.
In addition, a few points should be noticed.

First, at high metallicities, the abundances of La, Nd, and Sm with
respect to Fe are slightly overestimated as compared to spectroscopic
observations, whereas the ratios to Eu are well reproduced.  On the
other hand, for Ba, Ce, and Pr, the predicted abundance ratios with
respect to Fe are in good agreement with the observational data,
whereas the ratios with respect to Eu appears to be slightly
underestimated.  The origin of discrepancies in the [element/Fe] ratio
might be attributed to specific characteristics of the Galactic
evolution model (e.g. the Fe production by SNI/SNII). The ratio
[element/Eu], however, is much less model dependent, and discrepancies
between model results and observations should be attributed more likely
to incomplete observational data.

Second, europium deserves a particular attention as it is one of the
few $r$-process elements that has clean atomic lines accessible in the
visible part of the spectrum, which makes it an important diagnostic of
the $r$-process history of stellar material.  In Fig. 12 we show the
observationally determined values of [Eu/Fe] in a sample of halo and
disk stars. For [Fe/H] $\ge -2.4$, the similarity of O and Eu trends vs. [Fe/H] 
strongly supports the idea of a common production site,
presumably SNII.  In the same Fig. 12 we also show the results of our
chemical evolution model, obtained following the prescriptions
described in Sect.~4.2.  The scatter in [Eu/Fe] at lower metallicities
(also observed in the other Ba-peak elements) can be ascribed to an
incomplete mixing in the Galactic halo gas, allowing the formation of
stars super-rich in $r$-process elements, like CS 22892-052 (see Sneden
et al.~1996).  This star in particular, with [Fe/H] $\simeq -3.1$,
shows $r$-process enhancements of 40 times the solar value (e.g.
[Eu/Fe] $\simeq +1.7$, and [Ba/Fe] $\simeq +0.9$), much larger than the
abundances observed in any normal halo star. Nevertheless its [Ba/Eu]
is in good agreement with the typical $r$-process ratios we predict at
this metallicity.  Therefore CS 22892-052 is likely to be a
star born in an environment more strongly polluted by supernova debris
than the average halo gas; its peculiar abundances support the idea of
an inhomogeneous composition of the halo at early times.

Third, at low metallicity, the values of [element/Fe] shown in Fig.$6 -
11$ suggest an {\em intrinsic} scatter over about 2 dex.  McWilliam et
al. (1995) and McWilliam (1998) have argued that the observed
dispersion in heavy element abundances at low metallicity entirely
reflects the actual inhomogeneities in the chemical composition of the
gas from which these extremely metal-poor halo stars were formed.
Ryan, Norris, \& Beers (1996) took a step further and speculated that
the resulting chemical enrichment of the gas would be highly
inhomogeneous and basically limited to the supernova's sphere of
influence.  The problem of the observed spread in the measured stellar
abundances, in connection with mixing of heavy elements in the various
Galactic zones is a complicated one, and in our opinion not fully
understood.  The spread observed in the Galactic disk population by
Edvardsson et al.~(1993) may result either from the multizone structure
of the Galaxy (the thick and thin disk contribute in different ways to
the local enrichment), or from a contamination of contiguous radial
zones (see Moll\`a et al.~1997 for details). In the halo, the non
locality of observed populations will be again a good reason for a
relevant spread in the abundance distribution.  The question remains if
these mechanisms can be responsible for the extremely large spread in
the spectroscopic data presented by McWilliam~(1998).

We must keep in mind that models of Galactic chemical evolution often
assume instantaneous mixing between stellar ejecta and the ISM, and
they can therefore  provide only {\em spatially averaged} values of
element abundances as a function of time and galactocentric distance.
While such an assumption is justified during the evolution of the
Galactic disk and the late times of the halo phase, a different kind of
approach should be taken when investigating the earliest phases of halo
enrichment, where a coupled treatment of the chemistry and the dynamics
of the gas, both on a large scale (collapse) and on a small scale
(supernovae), is needed for a more realistic modeling.  Preliminary
results concerning Ba are shown in Cravanzola et al.~(1999), where the
chemo-dynamical evolution of the Galaxy is followed by means of a
N-body/SPH code. Similar conclusions have been advanced by Ishimaru \&
Wanajo~(1999).

On the other hand, the consistent observational scatter at very low
metallicities can be attributed, at least in part, to uncertainties in
the spectroscopic observations, which suffer for poor signal-to-noise
ratios due to the intrinsic weakness of the sources (Gratton \&
Sneden~1994; Gratton 1995).

The $s$-process contributions to the various isotopes of the elements
considered in this paper are listed in Table~3.  For example, at
$t=t_{\odot}$ the model can account for 94\% and 97\% of the $s$-only
isotopes $^{134}$Ba and $^{136}$Ba, respectively.  The dominantly
$r$-process Eu isotopes are affected by the $s$-process only at the 5\%
level. The solar distribution of $s$- and $r$-fractions of individual
isotopes is reproduced within $\sim 10$\%.

\begin{table}[t]
\begin{center}
TABLE 3\\
{\sc $s$-process fractional contributions (\%) for isotopes from Ba to Eu\\
at $t=t_{\odot}$ with respect to solar system abundances}\\
\vspace{0.5em}
\textheight 25cm
\textwidth 18cm
\voffset -2.5cm
\begin{tabular}{llllll}
\hline\hline
$^{134}$Ba  &  94 & $^{141}$Pr  & 47 & $^{147}$Sm  & 20   \\
$^{135}$Ba  &  22 & $^{142}$Nd  & 93 & $^{148}$Sm  & 96   \\
$^{136}$Ba  &  97 & $^{143}$Nd  & 30 & $^{149}$Sm  & 12   \\
$^{137}$Ba  &  58 & $^{144}$Nd  & 48 & $^{150}$Sm  & 94   \\
$^{138}$Ba  &  84 & $^{145}$Nd  & 26 & $^{152}$Sm  & 21   \\
$^{139}$La  &  61 & $^{146}$Nd  & 61 & $^{154}$Sm  & 0.5  \\
$^{140}$Ce  &  81 & $^{148}$Nd  & 13 & $^{151}$Eu  & 6    \\
$^{142}$Ce  &  12 & $^{150}$Nd  & 0.02  & $^{153}$Eu  & 5 \\
\hline \hline
\end{tabular}
\end{center}
\end{table}

\section{Conclusions}

In this paper we have calculated the evolution of $n$-capture elements
from Ba to Eu in the interstellar gas of the Galaxy.  The input stellar
yields for neutron-rich nuclei have been separated into their $s$- and
$r$-process components.  We have obtained the $s$-yields with
post-process calculations based on AGB models computed with the FRANEC
code.  The results of the Galactic evolution model, compared with
spectroscopic observations of F and G dwarf stars, confirm the basic
nucleosynthesis scenario outlined by Straniero et al.~(1997) and
Gallino et al.~(1998).  The $^{13}$C neutron source, active during the
interpulse phases of low-mass TP-AGB stars, accounts for most of the
$s$-process contribution to elements from Ba to Eu. In the Galactic
disk the abundance of Ba is dominated by the $s$-process for [Fe/H]
$\gtrsim - 1.5$. At lower metallicities, the primary $r$-process
contribution, relatively small at the time of formation of the Sun
(20\% of solar Ba), plays a dominant role.

Concerning the enrichment of the Galaxy in the $r$-process elements,
the comparison of the predictions of the Galactic chemical model with
the available spectroscopic data for Population II stars suggests a
production from SNII slightly delayed with respect to the main phase of
oxygen enrichment. The overall decline of [$r$/Fe] vs. [Fe/H] in the
most metal-poor stars can be explained if the $r$-process derives from
core-collapse SNe at the lowest stellar mass limit, around 8--10~\msb.

The Galactic evolution of the lighter $s$-process nuclei (e.g. Sr, Y,
Zr) as well as of Pb will be presented in a forthcoming paper.
Spectroscopic determinations of the abundance of these elements in
Galactic halo stars are currently under way (see e.g. Sneden et al.
1998 for recent HST data).  Concerning Pb, preliminary results on the
$s$-fraction at $t=t_\odot$ and on its sharp increase at low
metallicities, confirm the prediction by Gallino et al.~(1998) that the
production of $^{208}$Pb (the {\em strong component}) has to be
entirely attributed to $s$-process occurring in low-metallicity,
low-mass AGB stars.

\acknowledgements

We would like to thank C. Arlandini, A. Chieffi, M. Limongi and M.
Lugaro for useful discussions. We also thank S. Sandrelli and U. Penco
for kind assistance with the numerical calculations. The work of C.T. and
D.G. is supported in part by grant Cofin98-MURST at the Osservatorio Astrofisico di
Arcetri.  The work of R.G. and M.B. is supported by grant Cofin98-MURST
at the Dipartimento di Fisica Generale, Universit\`a di Torino, and
Osservatorio Astronomico di Pino Torinese.

\newpage

\plotone{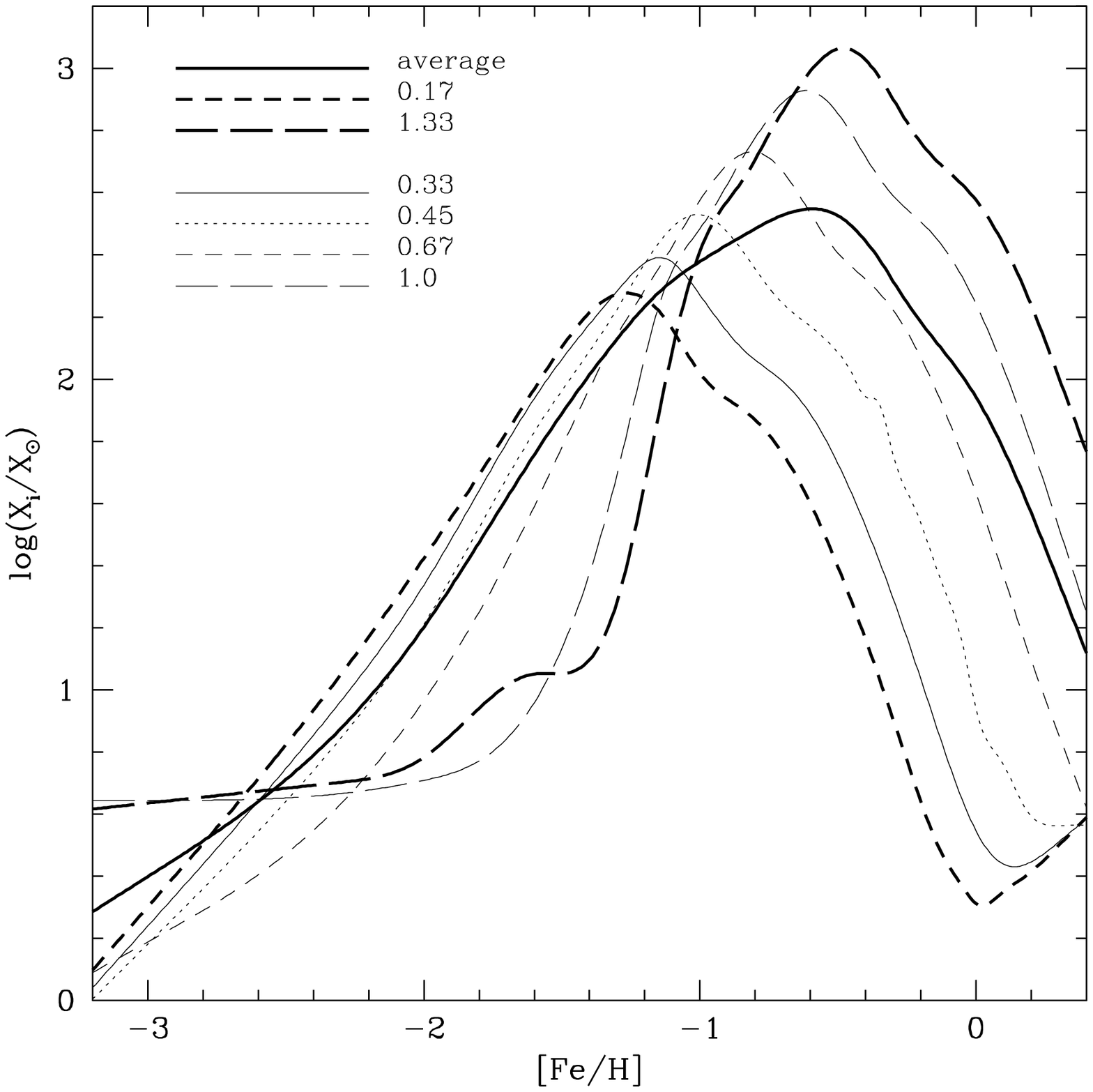}

\figcaption[fig1.eps]{Abundances by mass (relative to solar) of Ba in
the material cumulatively mixed to the surface of a 2~\msb star by
third dredge-up episodes as function of metallicity, for different
assumptions on the mass of the $^{13}$C pocket. The {\em short-} and
{\em long-dashed thick lines} represent the cases 0.17 and 1.33 times
the standard value (see text) respectively, whereas the {\em thin
lines} show intermediate cases (0.33, 0.45, 0.67, and 1 times the
standard value).  The {\em thick continuous line} represents the
unweighted average of all cases shown.\label{fig1}}

\plotone{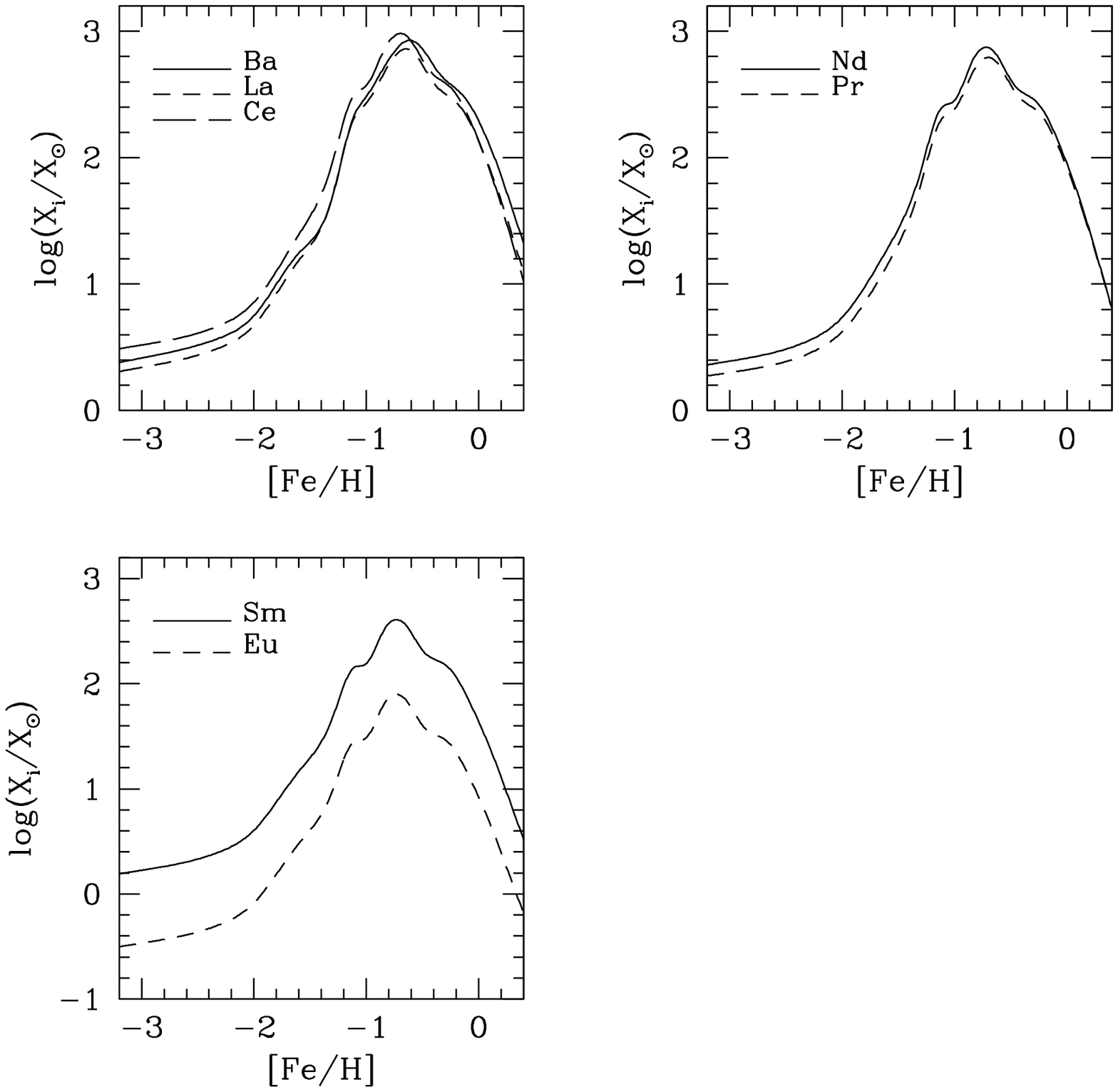}

\figcaption[fig2.eps]{Abundances by mass (relative to solar) of Ba, La,
and Ce ({\em top left panel\/}), Nd and Pr ({\em top right panel\/}),
Sm and Eu ({\em bottom left panel\/}) in the material  cumulatively
mixed to the surface of a 2~\msb star by third dredge-up episodes as
function of metallicity.\label{fig2}}

\plotone{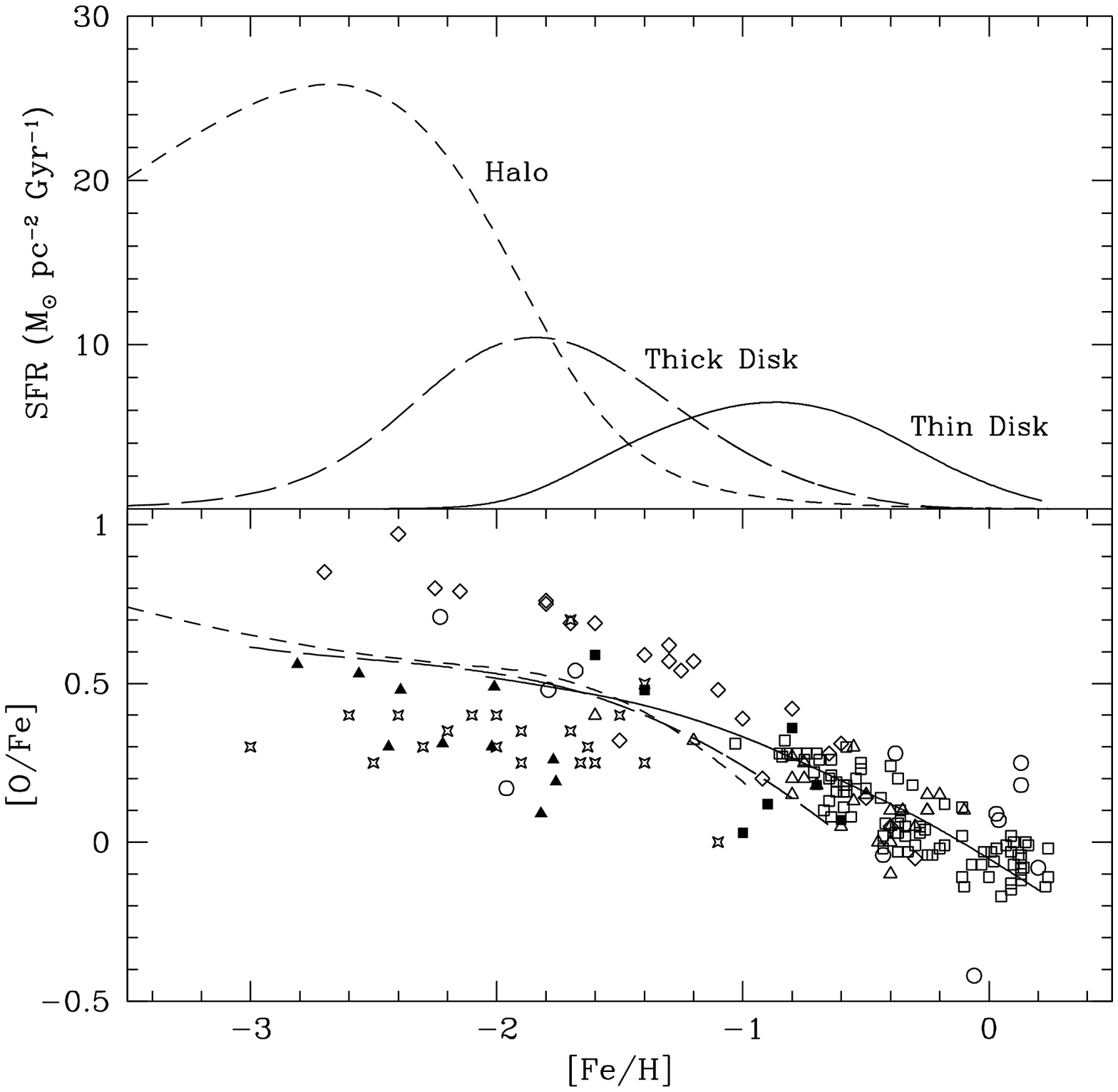}

\figcaption[fig3.eps]{Star formation rate ({\em upper panel\/}) and
[O/Fe] ({\em lower panel\/}) according to our standard model of
Galactic evolution, displayed as function of [Fe/H]. [O/Fe] predictions
are compared with spectroscopic observations by: Gratton \& Ortolani
(1986) ({\em circles}); Barbuy (1988) ({\em open triangles}); Barbuy \&
Erdelyi-Mendes (1989) ({\em four-pointed stars}); Sneden et al. (1991)
({\em filled triangles}); Spite \& Spite (1991) ({\em filled squares});
Edvardsson et al. (1993) ({\em open squares}); Israelian et al. (1998)
({\em diamonds}).\label{fig3}}

\plotone{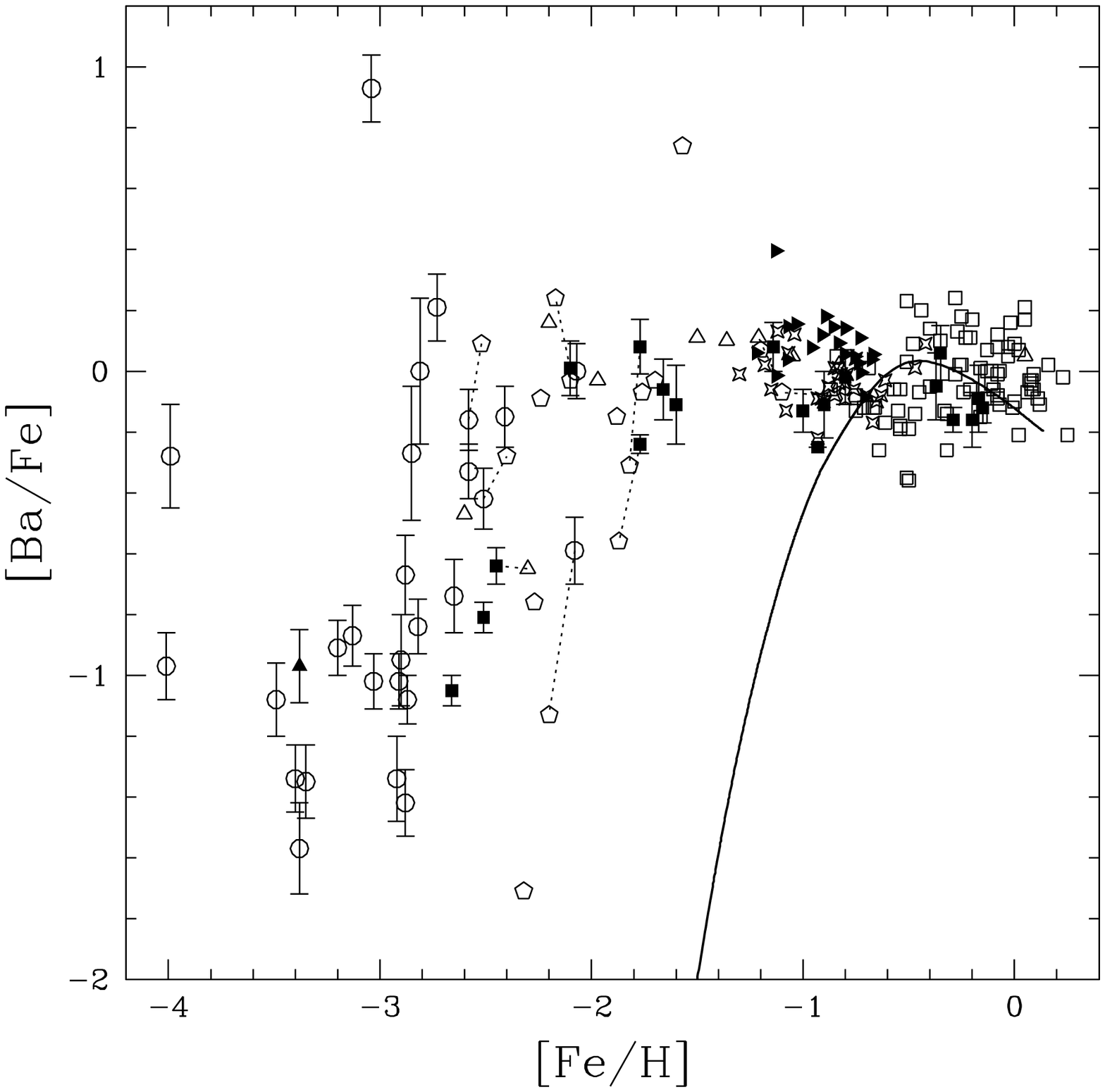}

\figcaption[fig4.eps]{Evolution of Ba $s$-fraction in the thin disk as
function of [Fe/H] according to our standard model ({\em solid line}).
Observational data are from Gratton \& Sneden~(1994) ({\em filled
squares}); Woolf et al.~(1995) ({\em open squares}); Fran\c cois~(1996)
({\em pentagons}); McWilliam et al.~(1995) and McWilliam~(1998) ({\em
circles}); Norris et al.~(1997) ({\em filled triangles}); Jehin et al.
(1998) ({\em filled tilted triangles}; Mashonkina et al.~1999 ({\em open
triangles}). Thin dotted lines connect stars
with different abundance determinations. Errorbars are shown only when
reported by the authors for single objects.  \label{fig4}}

\newpage

\plotone{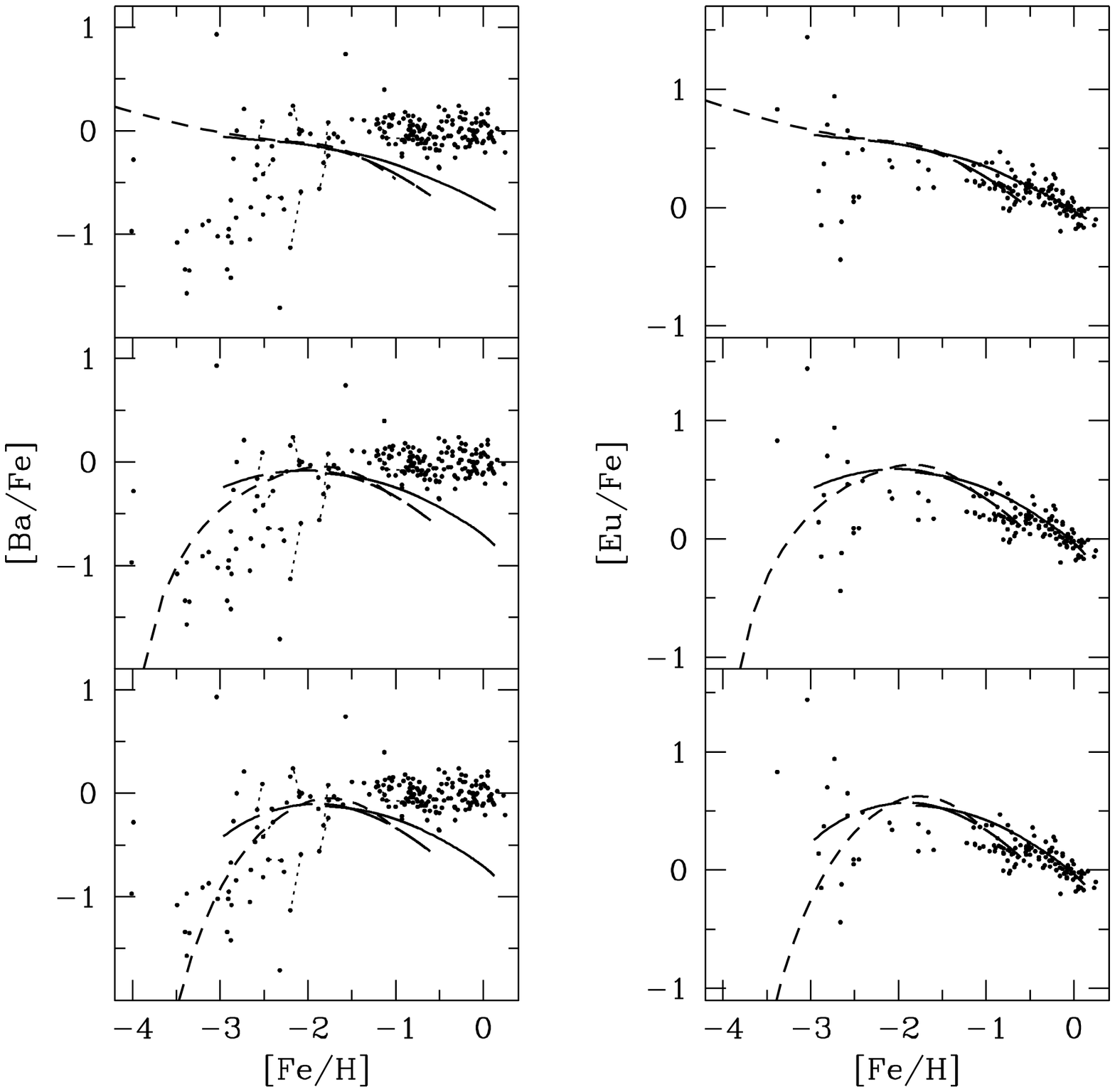}

\figcaption[fig5.eps]{Galactic evolution of the $r$-fraction of Ba
({\em left panels\/}) and Eu ({\em right panels\/}) according to our
standard model, for different assumptions on the mass range of SNII
contributing to the $r$-process nucleosynthesis (all SNII in the upper
panels, 10--12 $M_\odot$ SNII in the middle panels, 8--10 $M_\odot$
SNII in the bottom panels).  Lines refer to the halo ({\em
short-dashed}), thick-disk ({\em long-dashed}) and thin-disk ({\em
solid}). Data points are the same as in Fig.~4 and 12 (errorbars are
omitted for clarity).\label{fig5}}

\plotone{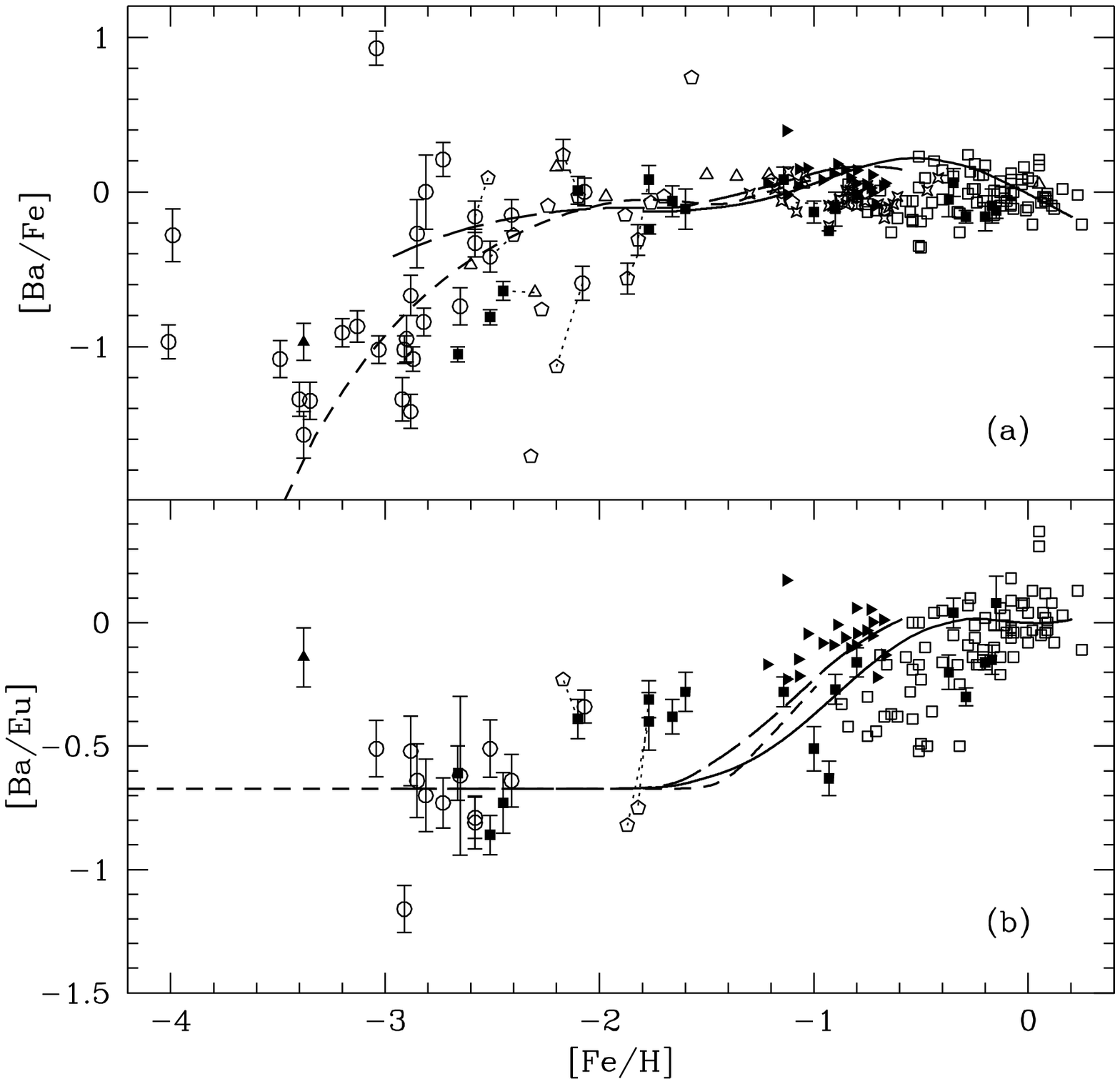}

\figcaption[fig6.eps]{Galactic evolution of [Ba/Fe] ({\em upper
panel\/}) and [Ba/Eu] ({\em lower panel\/}) according to our standard
model, including both the $s$- and $r$-process contributions. All
symbols are the same as in in Fig.~4. Errorbars are shown only when
reported by the authors for single objects.\label{fig6}}

\plotone{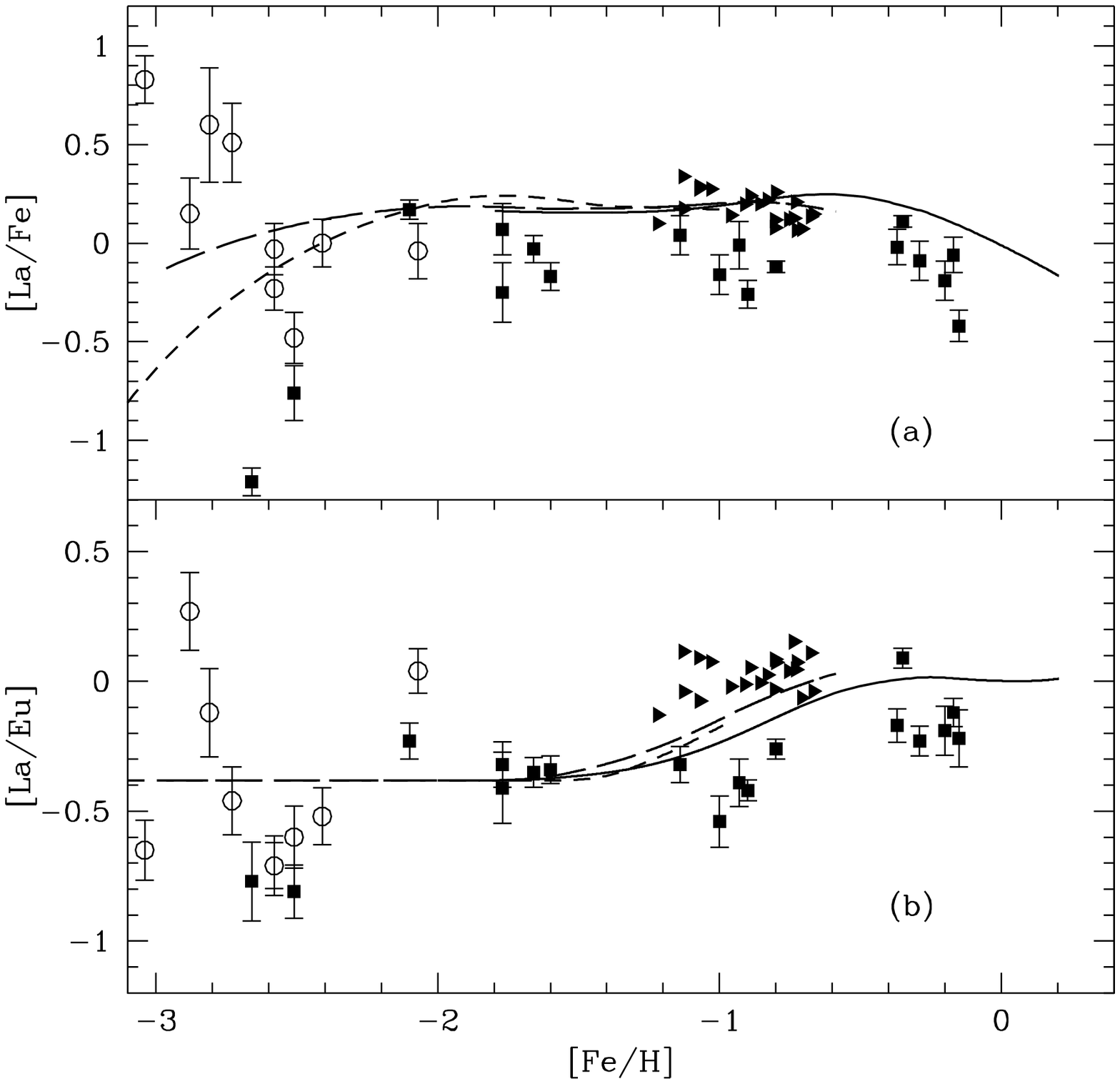}

\figcaption[fig7.eps]{Galactic evolution of [La/Fe] ({\em upper
panel\/}) and [La/Eu] ({\em lower panel\/}). All symbols are the same
as in in Fig.~4.  Errorbars are shown only when reported by the authors
for single objects.\label{fig7}}

\plotone{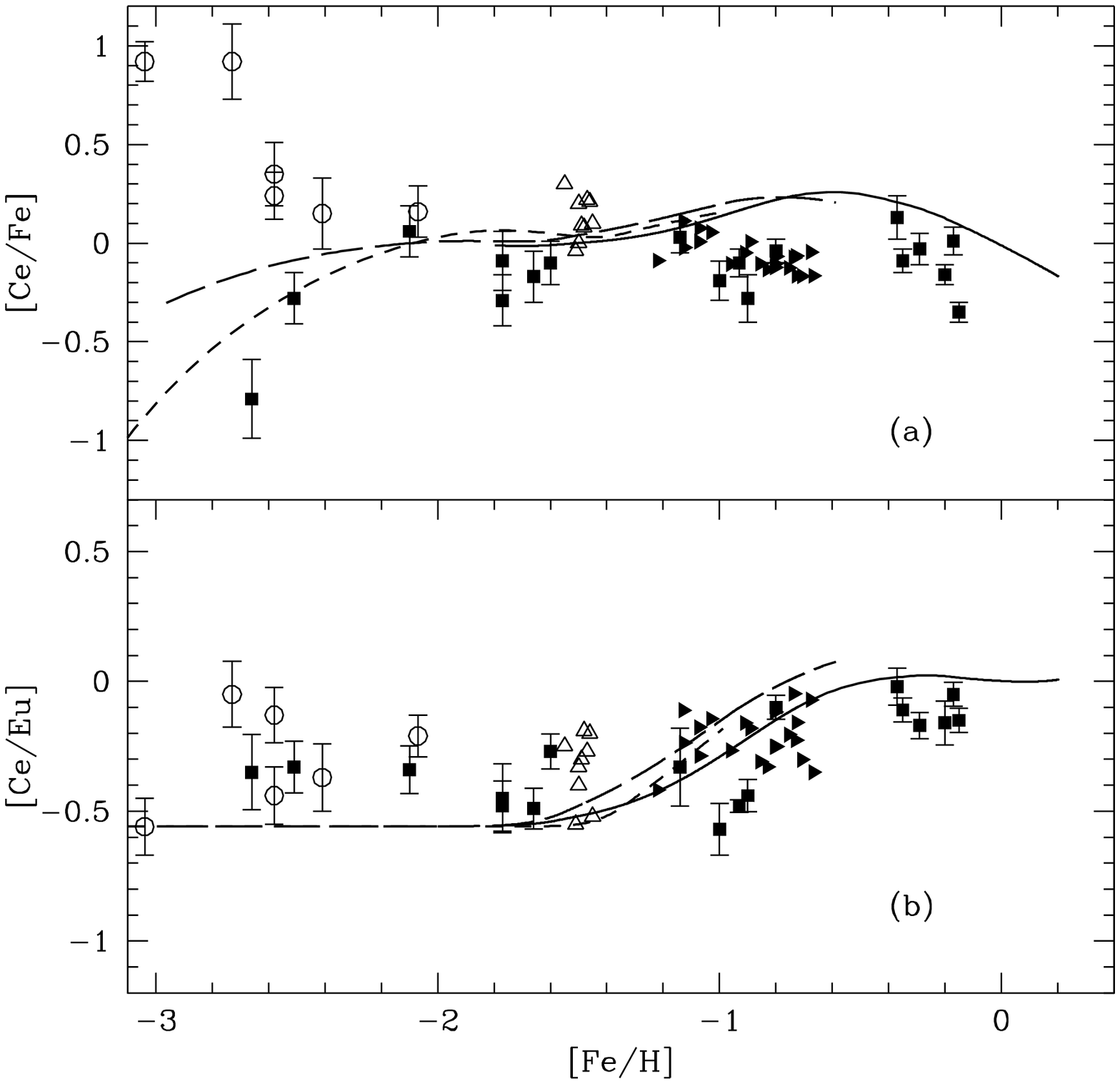}

\figcaption[fig8.eps]{Galactic evolution of [Ce/Fe] ({\em upper
panel\/}) and [Ce/Eu] ({\em lower panel\/}). All symbols are the same
as in in Fig.~4.  Errorbars are shown only when reported by the authors
for single objects.\label{fig8}}

\plotone{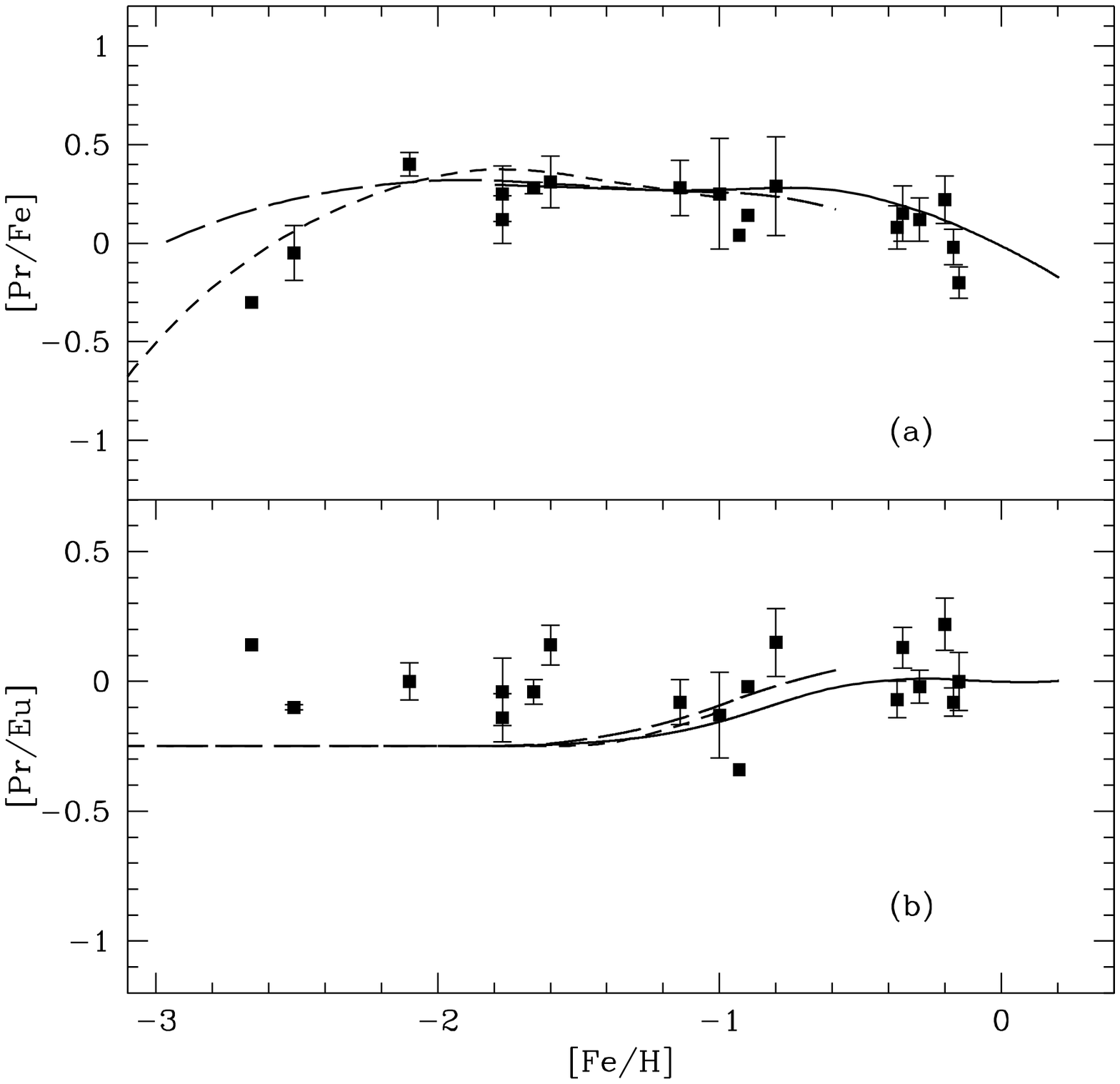}

\figcaption[fig9.eps]{Galactic evolution of [Pr/Fe] ({\em upper
panel\/}) and [Pr/Eu] ({\em lower panel\/}). All symbols are the same
as in in Fig.~4.  Errorbars are shown only when reported by the authors
for single objects.\label{fig9}}

\plotone{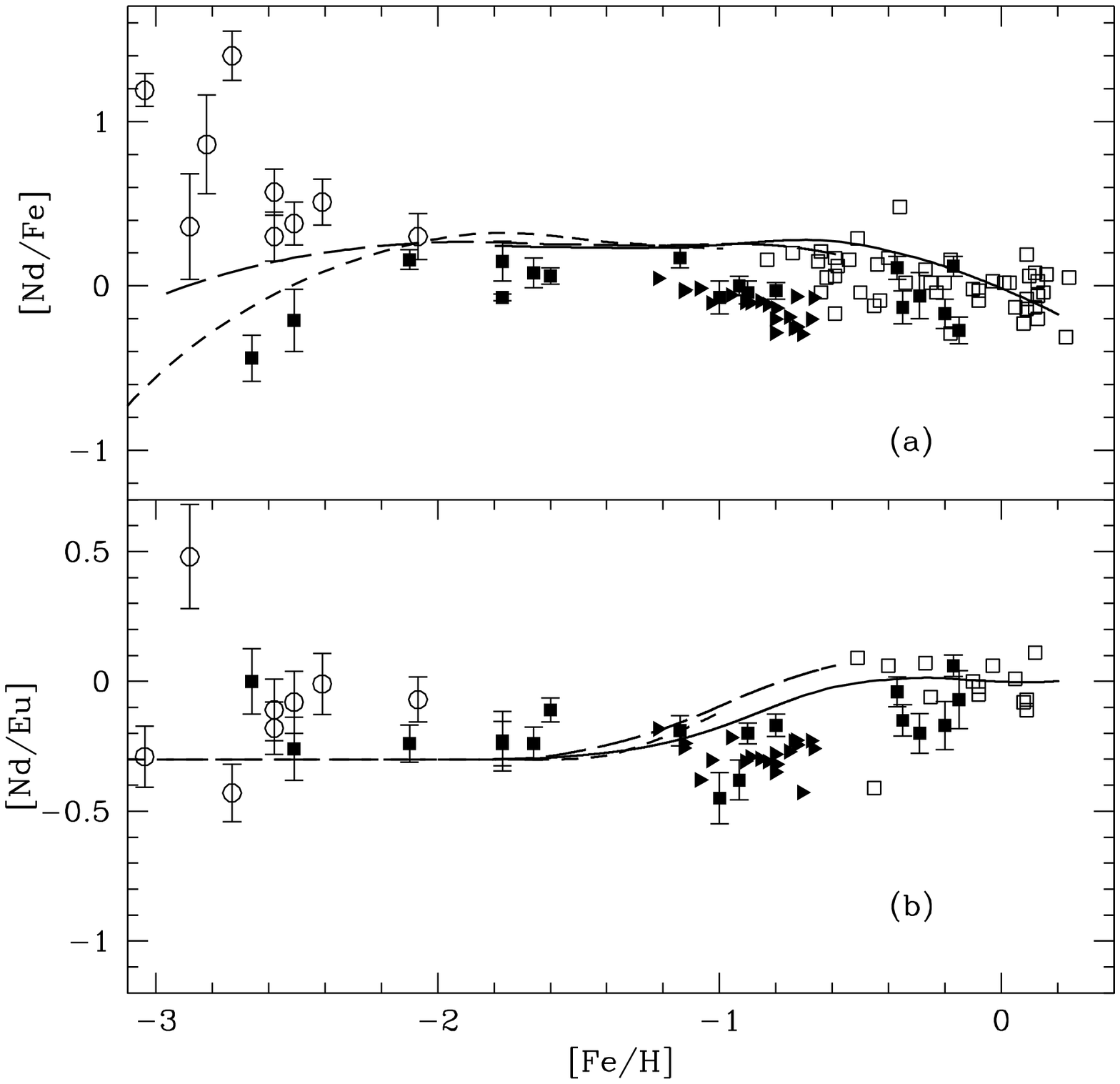}

\figcaption[fig10.eps]{Galactic evolution of [Nd/Fe] ({\em upper
panel\/}) and [Nd/Eu] ({\em lower panel\/}). All symbols are the same
as in in Fig.~4.  Errorbars are shown only when reported by the authors
for single objects.\label{fig10}}

\plotone{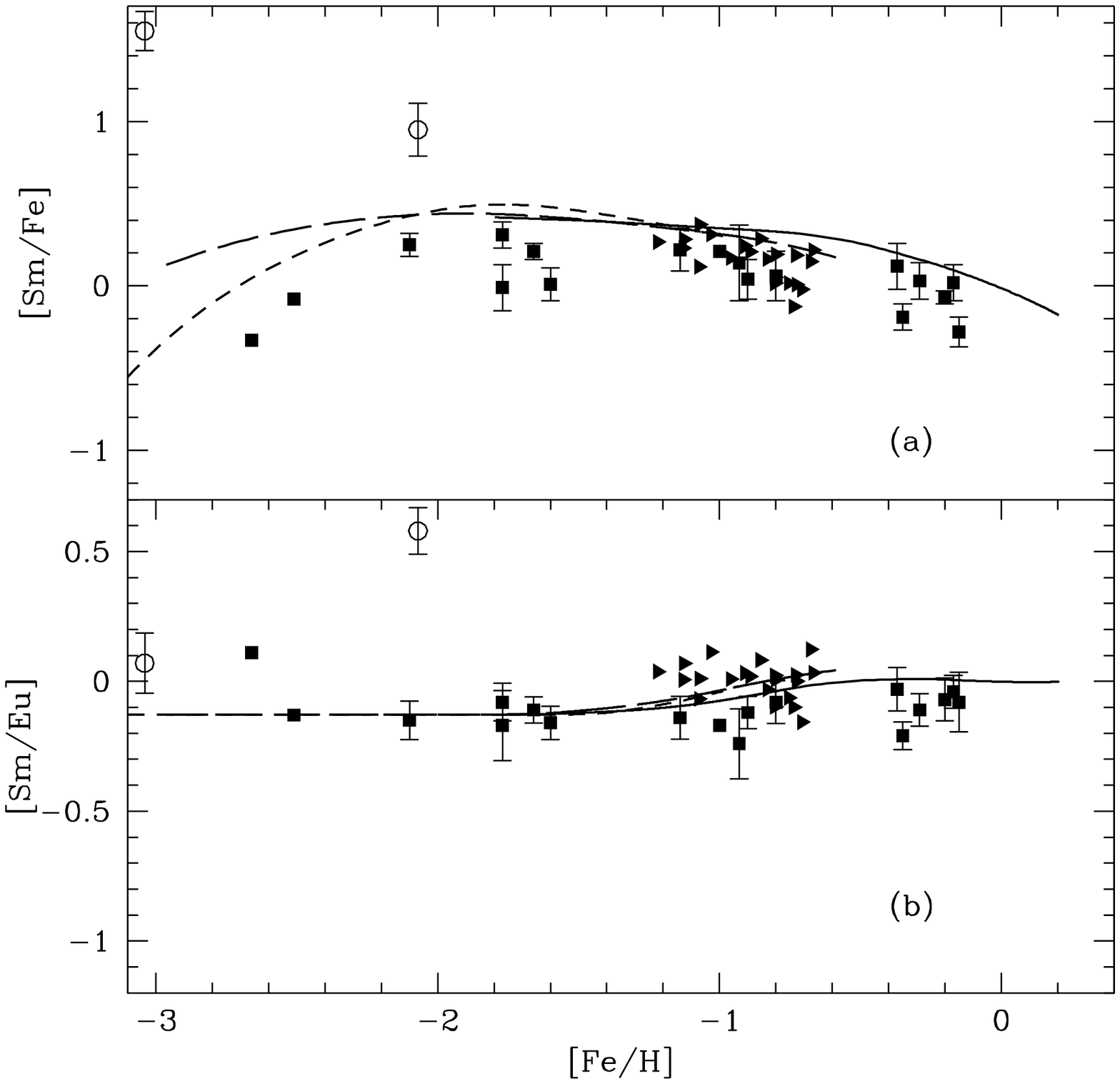}

\figcaption[fig11.eps]{Galactic evolution of [Sm/Fe] ({\em upper
panel\/}) and [Sm/Eu] ({\em lower panel\/}). All symbols are the same
as in in Fig.~4.  Errorbars are shown only when reported by the authors
for single objects.\label{fig11}}

\plotone{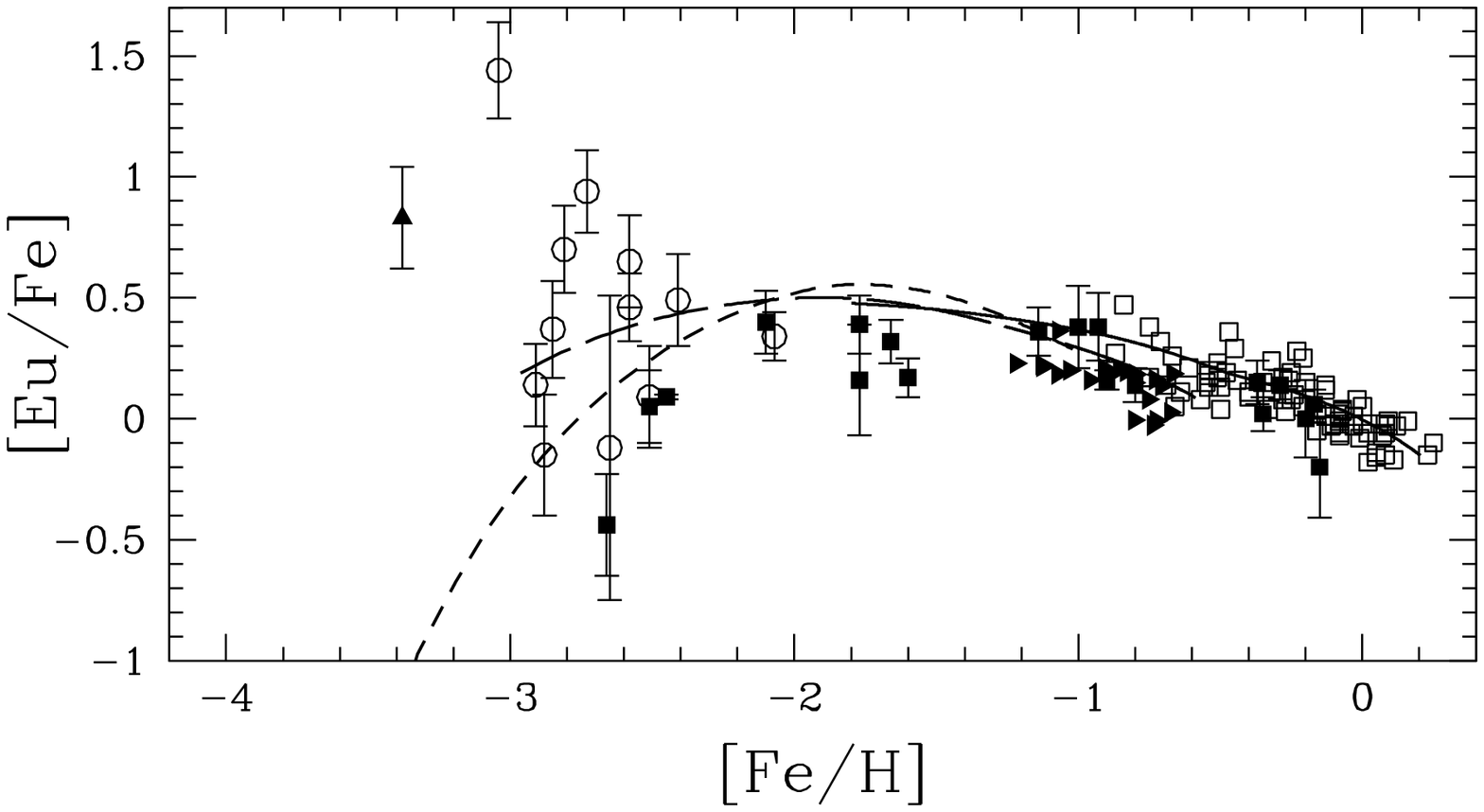}

\figcaption[fig12.eps]{Galactic evolution of [Eu/Fe] according to our
standard model, including both the $s$- and $r$-process contributions
(even at $t=t_{\rm Gal}$ the latter component accounts for 94\% of the
total production). All symbols are the same as in in Fig.~4. Errorbars
are shown only when reported by the authors for single
objects.\label{fig12}}

\plotone{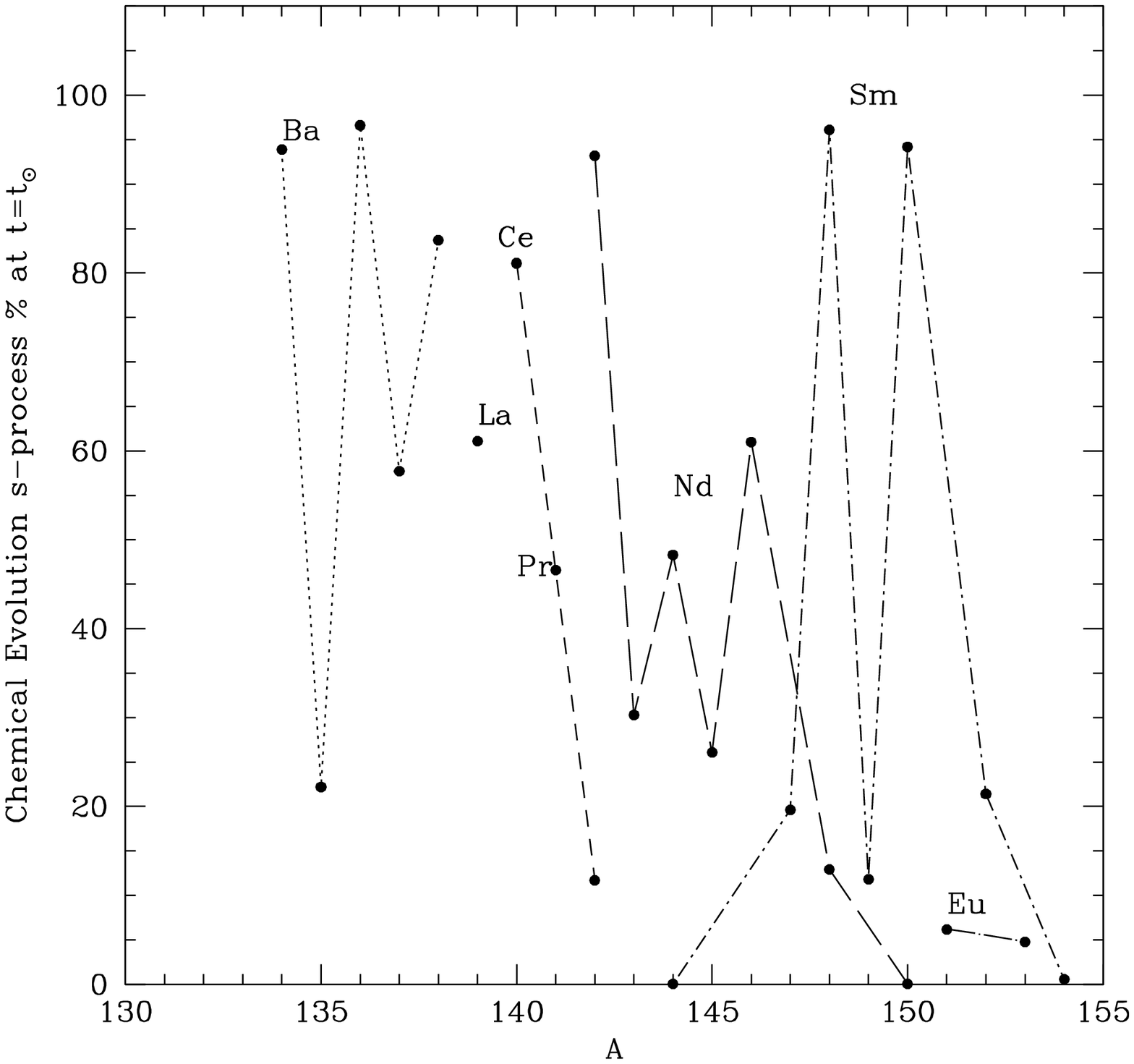}

\figcaption[fig13.eps]{$s$-fractions at $t=t_\odot$ of Ba to Eu
isotopes, according to our standard model. Values are given in
percentage with respect to solar abundances.\label{fig13}}

\end{document}